\documentclass[aps,prb,twocolumn]{revtex4}
\usepackage{graphicx}

\begin{document}
\bibliographystyle{plaint}
\title{The Spintronic Properties of Rare Earth Nitrides}
\author{C.M. Aerts,$^1$ P. Strange, $^1$ M. Horne,$^{1}$ W.\,M. Temmerman,$^2$ Z. Szotek,$^2$
 and A. Svane$^3$}
\affiliation{$^1$Department of Physics, Keele University, Staffordshire, ST5 5DY, England\\
$^2$Daresbury Laboratory, Daresbury, Warrington WA4 4AD, England\\
$^3$Department of Physics and Astronomy, University of Aarhus, DK-8000,
Aarhus C, Denmark. }
\date{\today}
%
\begin{abstract}

The electronic structure of the rare earth nitrides is studied
systematically using the {\it ab-initio} self-interaction corrected 
local-spin-density approximation (SIC-LSD). This approach allows both a 
localised description of the rare earth $f-$electrons and an itinerant 
description of the valence electrons. Localising different numbers of 
$f$-electrons on the rare earth atom corresponds to different valencies, 
and the total energies can be compared, providing a first-principles 
description of valence. 
CeN is found to be tetravalent while the remaining rare earth nitrides
are found to be trivalent.
We show that these materials have a broad range of
electronic properties including forming a new class of half-metallic magnets 
with high magnetic moments and are strong candidates for applications in 
spintronic and spin-filtering devices.  
\end{abstract}
\maketitle
%
\section{Introduction}
\label{sec:intro}

Rare earth (RE) materials have high magnetic moments and form a wide range of
magnetic structures. It is the occupancy of the highly localised $4f$-states
that determines the magnetic properties while the other electronic
properties are determined principally by the itinerant $s-d$
electrons.\cite{jens} The standard method of determining the electronic 
properties of magnetic materials is density functional theory with a local 
spin density approximation (LSD) for the exchange-correlation 
energy.\cite{Kohn+Vash} Traditionally this method has been applied to the 
$s-d$ electrons in rare earths while the $f$-electrons have been treated in 
an atomic model. Attempts to include the $4f$-electrons within the LSD have 
had only very limited success because the LSD is insufficient to describe 
the strong correlations experienced by the rare earth $f$-electrons. In 
recent years more advanced methods of electronic structure determination 
such as LSD plus Self-Interaction Corrections (SIC)\cite{ZP,brisbane} and 
LDA+U\cite{anis97} methods have sought to remedy this problem and have met with 
considerable success. 

Currently there is a great deal of interest in spintronics and spin-filtering 
from both the fundamental and applied point of view. There has been speculation 
in the literature that rare earth nitrides may form half-metallic 
ferromagnets.\cite{Birg71,Kald75,Hase77} This is surprising because, based on a 
simple ionic model, trivalent rare earth nitrides would be expected to be insulators with 
a similar electronic structure to the divalent rare earth chalcogenides. However 
if it is the case that they are half-metallic then they have potential 
applications in spintronic devices. To our knowledge, there has been 
no systematic study of the electronic structure of these materials at all. This 
is particularly surprising given that they all crystallise in the simple sodium 
chloride structure.\cite{Hull78} 

It is the purpose of this paper to discuss our calculations of the
electronic structure of rare earth nitrides and thus to gain a deeper
understanding of these materials and their potential applications. We have
performed SIC-LSD calculations for all the rare earth nitrides, in the 
ferromagnetic state, in both the divalent and trivalent states. Thus, in
this systematic study of rare earth nitrides we shall be concerned with the
nominal valence of the rare earth ion as we progress through the series.
Will these materials turn out to be all trivalent? What is the character of the 
rare earth $f$ and $d$ hybridization with the nitrogen $p$ states? Obviously,
the position of the empty $f$-states will be an important factor in determining
this. By performing this systematic study, we will be able to determine the
trends.

The remainder of the paper is organized as follows. After a brief summary of our 
calculational method in section II, in section III we discuss the detailed 
electronic structure of these materials, and trends in behaviour as one proceeds 
across the Periodic Table. We conclude the paper in section IV, indicating 
what our results imply for device applications.

\section{SIC-LSD}

The SIC-LSD approximation \cite{ZP} is an {\em ab-initio} electronic structure 
scheme, that is capable of describing localization phenomena in 
solids.\cite{brisbane,PRL1,Temmerman-NiO} In this scheme the spurious 
self-interaction of each occupied state, which is inherent in any local
approximation to Density Functional Theory (DFT),\cite{hohen65,jones89} 
is subtracted from the 
conventional LSD approximation to the total energy functional 
\begin{eqnarray}
E^{SIC}=E^{LSD}-\sum_{\alpha}\delta_{\alpha}^{SIC},
\end{eqnarray}
where $\alpha$ labels the occupied states and $\delta_{\alpha}^{SIC}$ is
the self-interaction correction for state $\alpha$. This leads to a greatly 
improved description of static Coulomb correlation effects over the LSD 
approximation. Demonstrations of the advantages of the SIC approach have 
been made in studies of the Hubbard model,\cite{Rapid,vogl-hub} in 
applications to 3d oxides such as monoxides,\cite{PRL1,Temmerman-NiO} 
high temperature cuprates,\cite{wmt-YBCO,PRL2,Temmerman-La2} 
and magnetite,\cite{fe3o4} to $f$-electron systems,\cite{nature,selen,science} and to solid 
hydrogen.\cite{SSC}


In the SIC-LSD method it is necessary to minimise the total energy with
respect to the number of localised electrons which also leads to a
determination of the nominal valence defined as the integer number of
electrons available for band formation
$$N_{v}=Z-N_{core}-N_{SIC}$$
where $Z$ is the atomic number of the rare earth and $N_{core}$ is the 
number of atomic core electrons. $N_{SIC}$ is the number of $f$-states for 
which the self-interaction correction has been removed and is determined 
so that $N_v$ equals 3 for trivalent, and 2 for divalent, systems. 

Further details of the SIC-LSD method are discussed elsewhere.\cite{brisbane}

\section{Results and Discussion}

\subsection{Energetic Properties}

\begin{figure}
\includegraphics[scale=0.45, angle=0]{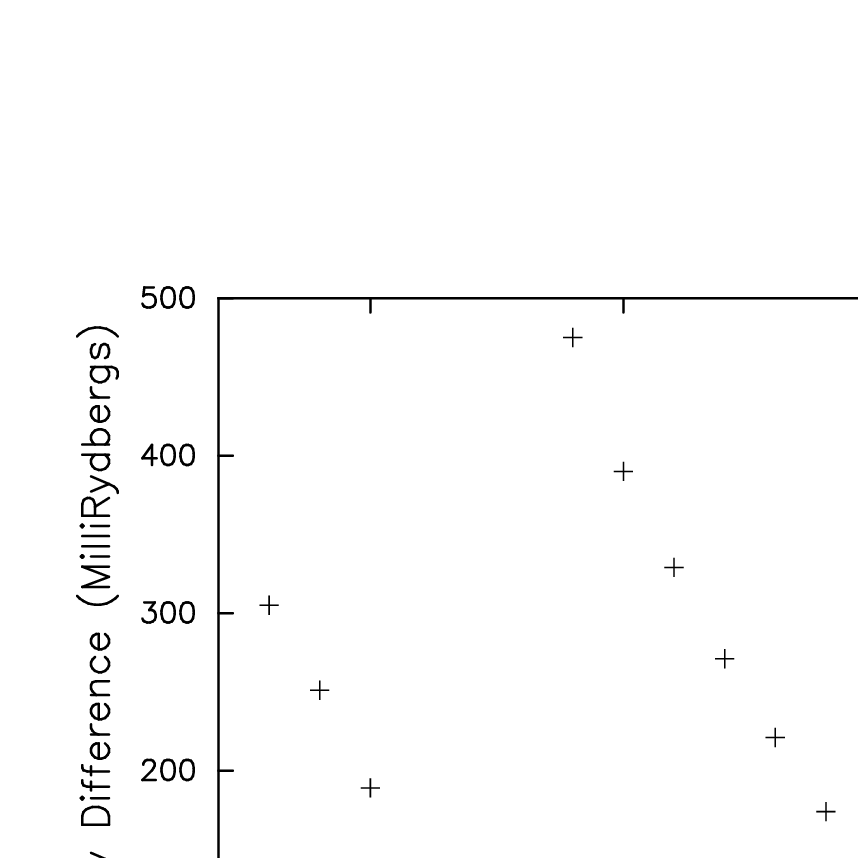} 
\vskip 2cm
\caption{The difference in total energy between the divalent and trivalent
states of the rare earth nitrides. The differences are all positive
indicating that the trivalent state is preferred over the divalent state in 
all cases in agreement with the published literature. Unlike in the previous
publications\cite{yb12} no spin-orbit coupling has been included in the present calculations.}
\label{fig1}
\end{figure}

\begin{figure}
\includegraphics[scale=.45,angle=0]{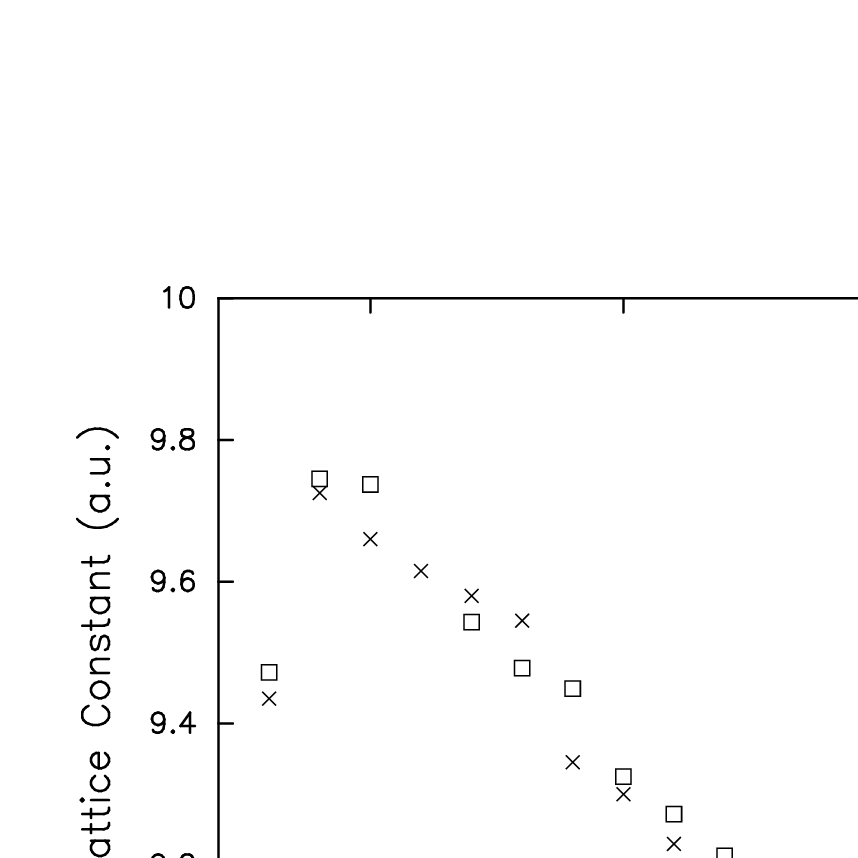}
\vskip 2cm
\caption{The lattice constant for the rare earth nitrides, crosses are the
calculated values and the squares are experimentally determined for thin
films of the material.\cite{Wyck}} 
\label{fig2}
\end{figure}

We have performed calculations for all the rare earth nitrides in 
ferromagnetic ordering and both the 
divalent $f^{n+1}$ and trivalent $f^n$ configurations (and for CeN the 
tetravalent $f^{n-1}$ configuration), and hence have determined the ground 
state energy and valence from first principles.
We have found that the ground state for Ce in CeN is tetravalent while all 
the other rare earths are trivalent in the nitrides. In Fig. \ref{fig1}, we show 
the difference in energy between the divalent and trivalent states of all 
the rare earth nitrides. This follows the expected trend of strongly 
trivalent state at the start of the series in both spin channels, and a 
decreasing energy difference between the trivalent and divalent state as 
the filling of the $f$-shell occurs. However, none of the nitrides attains 
the divalent state. Furthermore the number of occupied itinerant $f$-electron 
states stays well below the 0.7 required for the materials to become 
divalent, as observed by Strange {\it et al.},\cite{nature} because of the
strong hybridisation with the nitrogen $p$ states.
The lattice constants for the $f$ configuration that yields the lowest total
energy can be evaluated by looking for the minimum in total energy as a 
function of lattice size and these are shown in Fig. \ref{fig2} along with the 
experimental values.\cite{Wyck} The jump in lattice constant between Ce and 
Pr is due to the valence change from tetravalent to trivalent. For Ce, Pr 
and the heavy rare earths, theory and experiment agree very well both for 
the absolute values and the trend. For the remaining light rare earths the 
absolute agreement is still very good but there is a trend apparent in the 
theory that is not clear in the experimental results. The most probable 
cause of this is experimental uncertainty which can arise in these materials 
very easily, particularly as many of these measurements were performed at a 
time when samples of rare earth materials were known to have very limited 
purity.\cite{Wyck}  

\subsection{Magnetic Properties}

Now let us look at how the magnetism varies as we proceed along the rare
earth series. The total and species-decomposed spin magnetic moments are 
displayed in Table \ref{table1}. There we also present the rare earth's
orbital moments. It should be noted that there is some arbitrariness 
involved in the values of the spin magnetic moments due to the Atomic Sphere
Approximation. However, we found this to be small and moreover to make
comparisons between different systems consistent we kept the ratio of the
rare earth/nitrogen sphere volumes constant. 
\begin{table}
\caption{Various components of the spin magnetic moment (M$_{S}$) of the 
rare earth nitrides. All values are in Bohr magnetons ($\mu_{B}$). Where 
the numbers do not quite add up to the total the remaining spin moment is 
in the empty spheres used in the calculation. The rare earth orbital moment 
(M$_{L}$) assumes that the rare earth ions obey Hund's rules.}
\begin{tabular}{lcccccccc}
Material  & &       &  & M$_{S}$ &  &         & & M$_{L}$  \\
          & & {RE}  &  & {N}     &  & {Total} & & {RE}     \\
\hline
CeN & & 0.0  & &  0.0  & & 0.0  & &  0.0  \\ 
PrN & & 2.07 & & -0.08 & & 2.00 & & -5.0\\ 
NdN & & 3.10 & & -0.11 & & 3.00 & & -6.0 \\ 
PmN & & 4.13 & & -0.14 & & 4.00 & & -6.0 \\ 
SmN & & 5.22 & & -0.24 & & 5.00 & & -5.0 \\ 
EuN & & 6.30 & & -0.30 & & 6.00 & & -3.0 \\ 
GdN & & 7.01 & & -0.04 & & 7.00 & & -0.0 \\ 
TbN & & 5.97 & &  0.01 & & 6.00 & & 3.0 \\ 
DyN & & 4.93 & &  0.05 & & 5.00 & & 5.0 \\ 
HoN & & 3.91 & &  0.08 & & 4.00 & & 6.0 \\ 
ErN & & 2.90 & &  0.09 & & 2.99 & & 6.0 \\ 
TmN & & 1.83 & &  0.12 & & 1.96 & & 5.0 \\ 
YbN & & 0.79 & &  0.14 & & 0.94 & & 3.0 \\ 
\end{tabular}
\label{table1}
\end{table}

With the exceptions of ErN, TmN and YbN, the spin magnetic moments of the
rare earth nitrides 
take on an integer value. This indicates that these systems are
either insulating (Tb-, Dy-, and Ho-nitrides) or half-metallic (Pr- to 
Gd-compounds). CeN is a non-magnetic metal, and the last three compounds of the
series are metallic in both spin-channels. These results are as one would expect; 
the spin magnetic moment is dominated by the rare earth $f$-electrons, with some 
hybridisation yielding small contributions from the rare earth $s-d$ electrons 
and the nitrogen $p$ states. This indicates that the nitrogen $p$ states occur
in the same energy range as the valence rare earth states, allowing
hybridization to occur. It is interesting to note that the contribution 
from the nitrogen atom changes sign half way through the series, in which it
follows the RE's orbital moment. It appears 
that the nitrogen moment wants to point antiparallel to the partially 
occupied $f$-spin channel. An explanation of this hybridisation phenomenon
will be given in sections III D and E. 

\subsection{Density of States}

\begin{figure}
\includegraphics[scale=.37,angle=0]{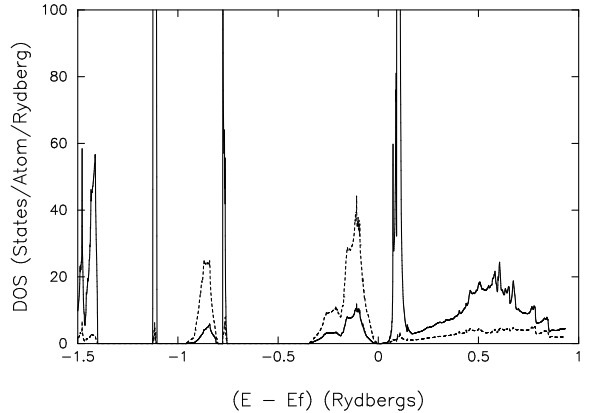}
\vskip 2cm
\caption{The density of states for Terbium Nitride in the trivalent state, 
(Tb Site = full line,  Nitrogen site = dashed line).}
\label{fig3}
\end{figure}

In Fig. \ref{fig3} we show typical densities of states of RE nitrides to illustrate 
the main features. Around -1.5 Ryd are the rare earth 5p bands. Above these, just 
above -1.0 Ryd, are the nitrogen 2s bands. Just below the Fermi energy $E_f$, we 
reach bands which are predominantly nitrogen $p$-like, but with a substantial 
hybridisation with rare earth $s-d$ and $f$-states. There is a small gap or 
mimimum around the Fermi energy. Above $E_f$ are the mainly rare earth $s-d$ 
bands. Superimposed on this are the rare earth $f$-bands. The fully occupied 
majority spin $f$-bands fall below the nitrogen 2$s$ bands and push the rare 
earth 5$p$ bands lower in energy. The occupied minority spin $f$-bands sit just 
above the nitrogen 2$s$ bands and there is essentially a single peak above E$_{f}$ 
which are the unoccupied $f$-states. Below half filling the occupied majority 
spin peak moves to just above the nitrogen 2$s$ bands, and just above $E_f$ 
there are two $f$-peaks, firstly the unoccupied majority spin bands and then 
the completely unoccupied minority spin $f$-bands. 

Let us look more closely at the electronic properties of the rare earth
nitrides. Table \ref{table2} shows the electronic properties of each material at the
Fermi energy.
\begin{table}

\caption{Spin resolved band gaps and densities of states at the Fermi 
energy for the rare earth nitrides. Band gaps are in Rydbergs and 
densities of states are in states/Rydberg/formula unit.}
\begin{tabular}{l c c c c}
Material & {Band} & {Gap} & {DOS} & {DOS} \\
         & {Spin-up} & {Spin-down} & {Spin-up} & {Spin-down} \\
\hline 
CeN & 0  & 0  & 15.37 & 15.37 \\ 
PrN & 0 & 0.039 & 0.0001 & 0 \\ 
NdN & 0 & 0.065 & 0.068 & 0 \\ 
PmN & 0 & 0.076 & 31.77 & 0 \\ 
SmN & 0 & 0.095 & 154.56 & 0 \\ 
EuN & 0 & 0.107 & 69.85 & 0 \\ 
GdN & 0 & 0.082 & 0.065 & 0  \\ 
TbN & 0.008 & 0.052 & 0 & 0 \\ 
DyN & 0.018 & 0.058 & 0 & 0 \\ 
HoN & 0.031 & 0.004 & 0 & 0 \\ 
ErN & 0 & 0  & 0.682  & 69.57 \\ 
TmN & 0 & 0 & 1.52 & 220.78 \\ 
YbN & 0 & 0 & 1.72 & 93.18 \\ 
\end{tabular}
\label{table2}
\vskip 2mm
\end{table}
It is clear from the table that most of the light rare earth nitrides are 
found to be half-metallic. Only CeN is metallic because it exists in the
tetravalent state (in fact the trivalent state is also just metallic). 
TbN, DyN and HoN are found to be narrow gap insulators and ErN, TmN and 
YbN are metallic in both spin-channels.  
\begin{table}
\caption{The spin resolved and $\ell$-decomposed densities of states at the Fermi 
energy for the rare earth nitrides in states/Rydberg/formula unit.}
\begin{tabular}{l c c c c c c}
Material &      & {DOS-up} &      &   & {DOS-down} &   \\
 & {RE s-d} & {RE f} & {N p} & {RE s-d} & {RE f} & {N p}  \\
& & & & & & \\
\hline 
CeN & 2.10 & 11.19 & 0.13& 2.10 & 11.18 & 0.13 \\ 
PrN & 0.0001 & 0 & 0 & 0 & 0 & 0  \\ 
NdN & 0.020 & 0.003 & 0.030 & 0 & 0 & 0 \\ 
PmN & 0.061 & 31.02 & 0.40 & 0 & 0 & 0 \\ 
SmN & 0.44 & 151.42 & 2.06 & 0 & 0 & 0 \\ 
EuN & 0.64 & 64.76 & 3.24 & 0 & 0 & 0 \\ 
GdN & 0.013 & 0.002 & 0.040 & 0 & 0 & 0  \\ 
TbN & 0 & 0 & 0 & 0 & 0 & 0 \\ 
DyN & 0 & 0 & 0 & 0 & 0 & 0 \\ 
HoN & 0 & 0 & 0 & 0 & 0 & 0 \\ 
ErN & 0.017  & 0.037 & 0.605 & 0.069 & 68.45 & 0.634 \\ 
TmN & 0.029 & 0.070 & 1.30 & 0.28 & 218.81 & 1.26 \\ 
YbN & 0.098 & 0.08 & 1.46 & 0.37 & 89.71 & 2.10 \\ 
\end{tabular}
\label{table3}
\vskip 2mm
\end{table}

In Table \ref{table3} we show the key spin- and $l-$decomposed densities of 
states at the Fermi energy. This table shows that the principal character of the
electrons at $E_f$ is rare earth $s-d$ and nitrogen $p$-like in
PrN, NdN, and GdN, while it is dominated by the majority spin rare earth  
$f$-electrons in PmN, SmN, EuN, and the minority spin rare earth
$f$-electrons in ErN, TmN and YbN. 
When the $f$ contribution is large we also note a relatively large N p
component, reflecting the strong N $p$ - rare earth $f$ hybridization.
From the results above it is also clear that 
there is a wide range of electronic properties in the rare earth nitrides 
which we discuss in detail in the following section.

\begin{figure*}
\begin{tabular}{l c c c c c c c c c c c c c c r}
\includegraphics[scale=.28, angle=0]{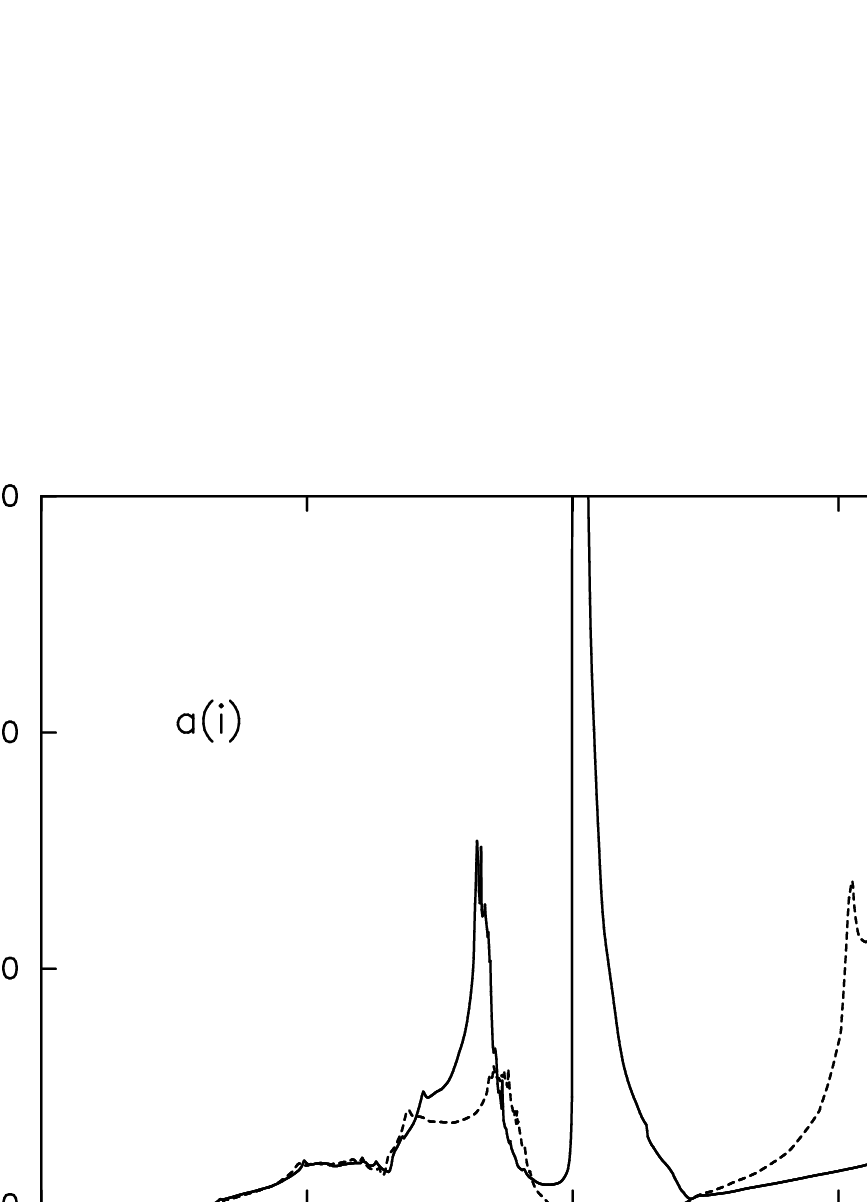} & & & & & & & & & & & & & & &  
\includegraphics[scale=.28, angle=0]{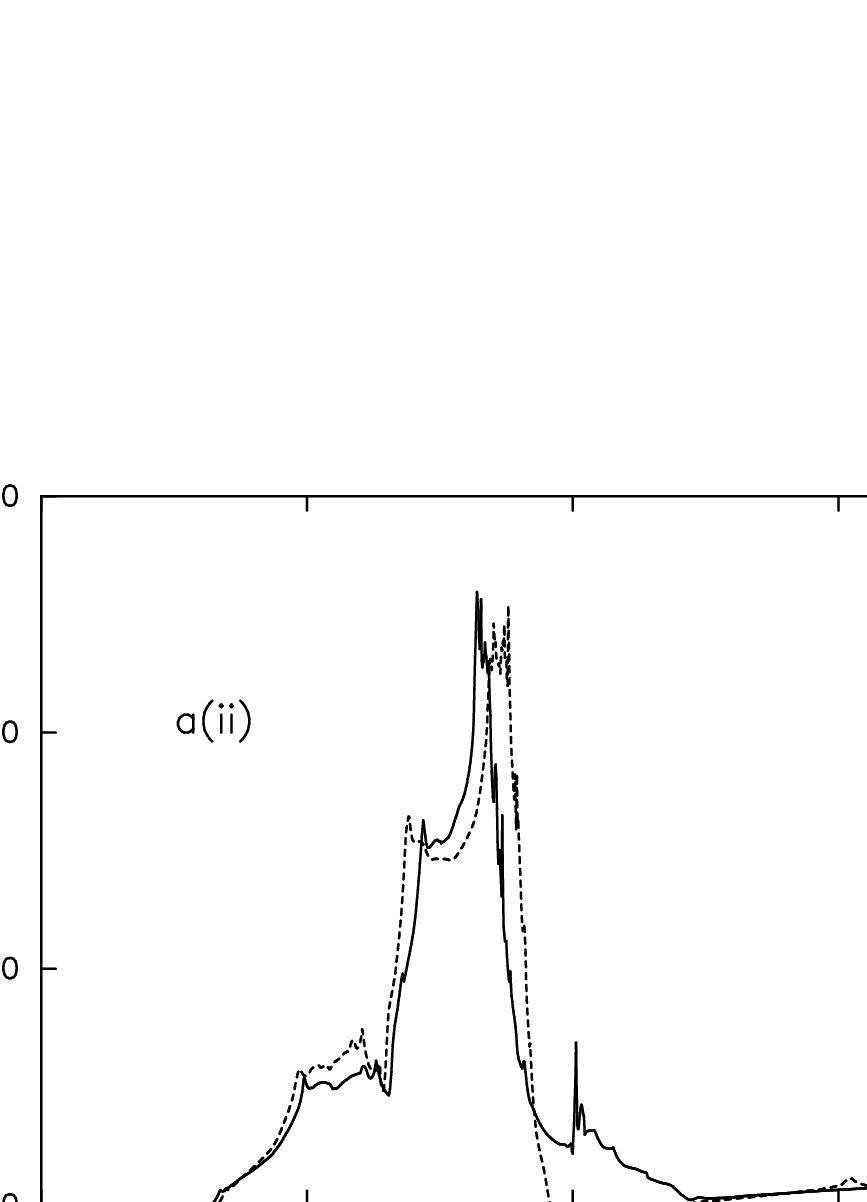} \\
\includegraphics[scale=.28, angle=0]{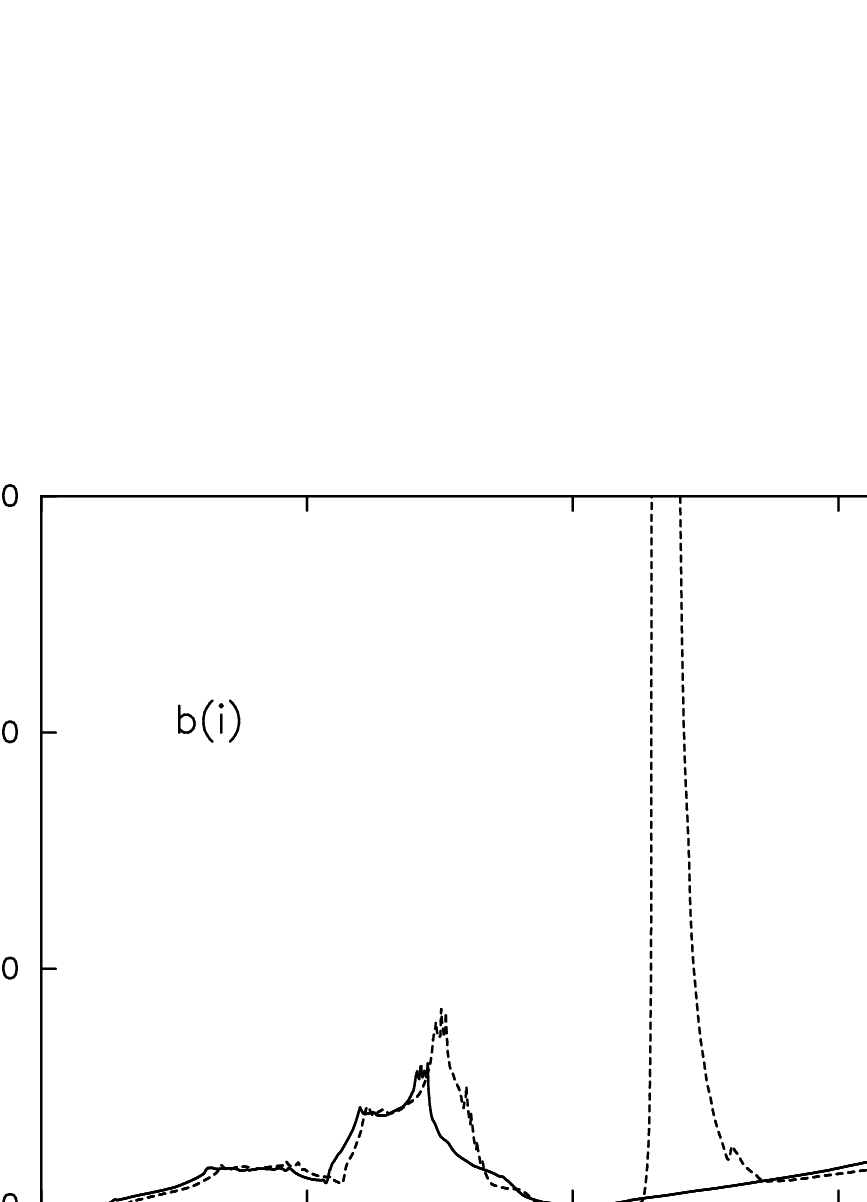} & & & & & & & & & & & & & & &
\includegraphics[scale=.28, angle=0]{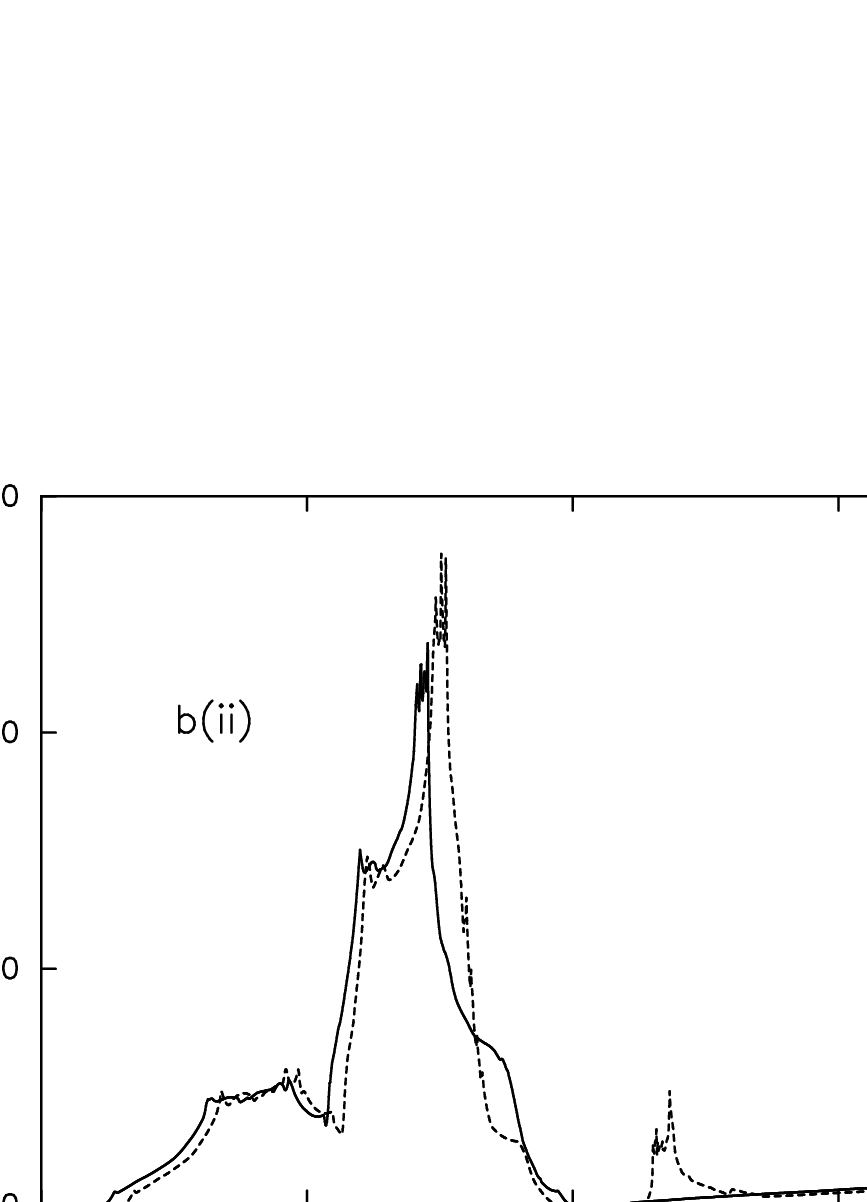} \\
\includegraphics[scale=.28, angle=0]{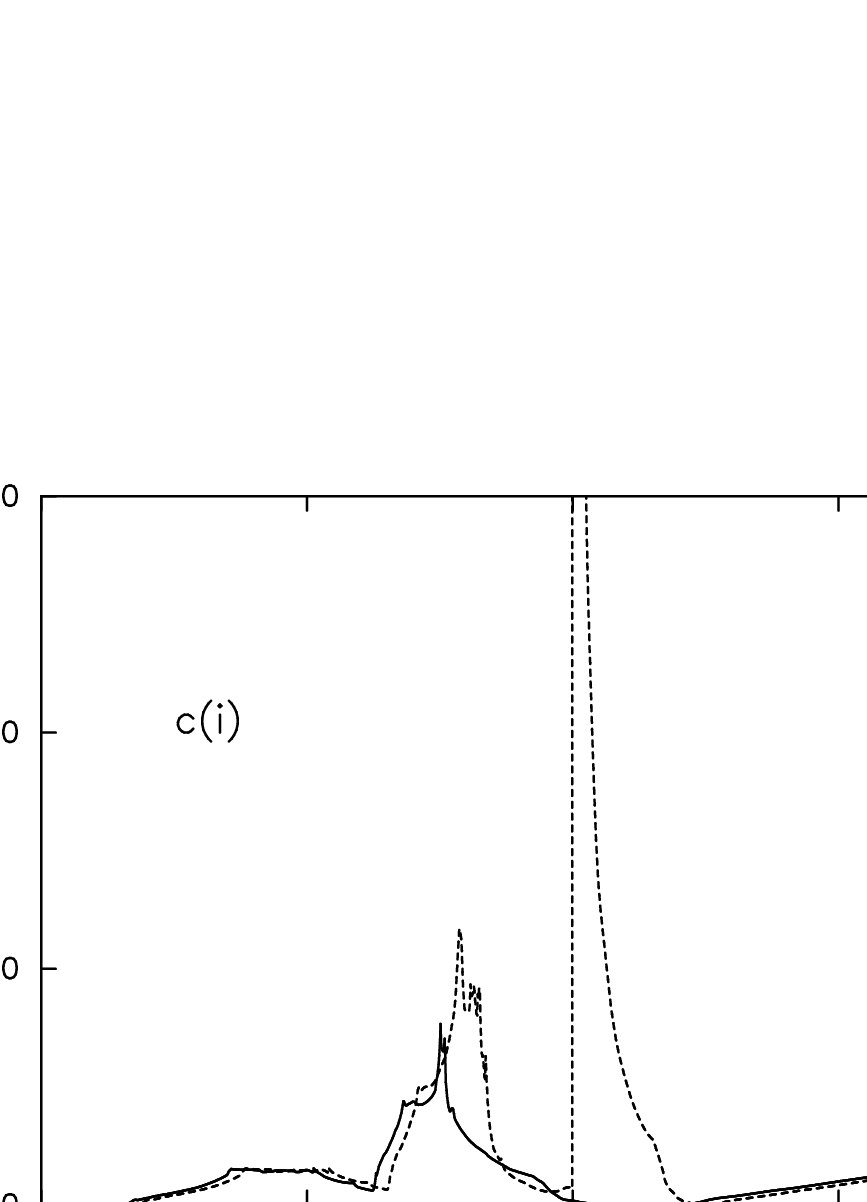} & & & & & & & & & & & & & & &
\includegraphics[scale=.28, angle=0]{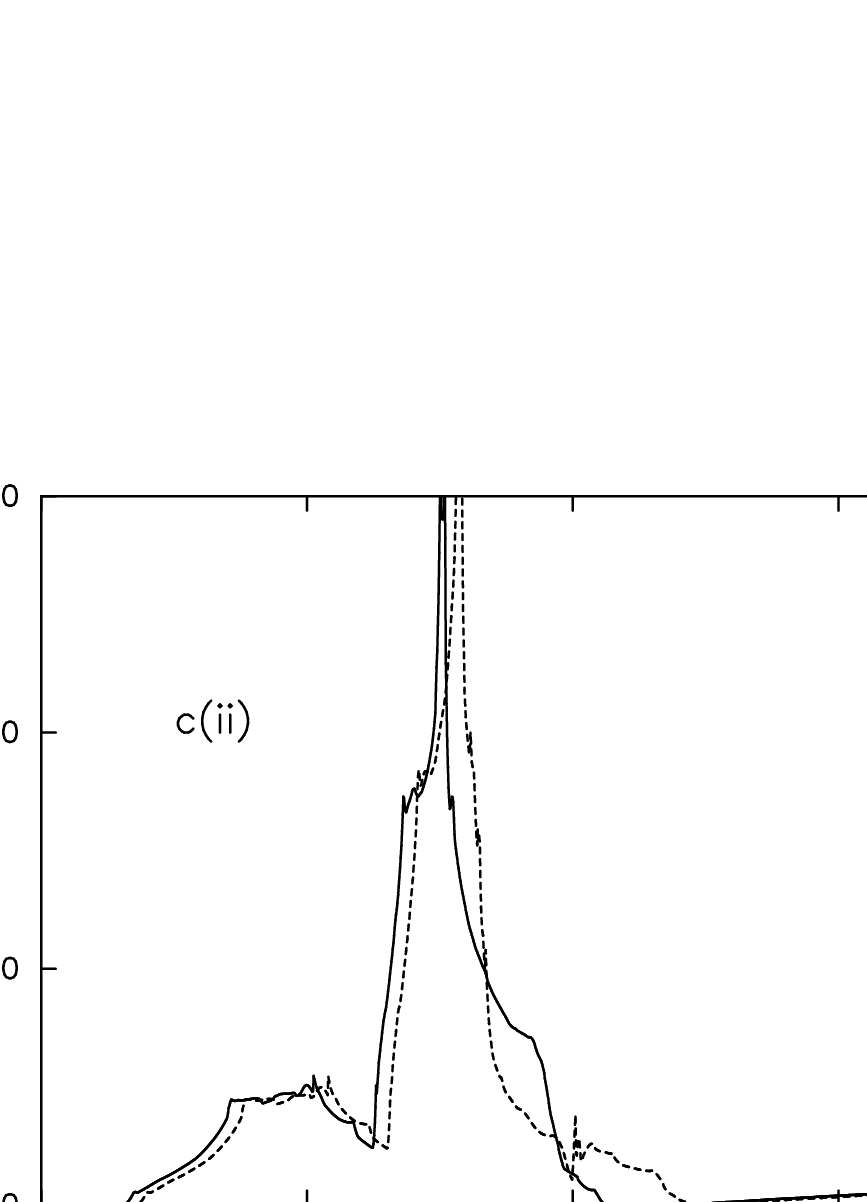} \\
\end{tabular}
\vskip 2cm
\caption{The density of states around the Fermi energy for (a) SmN, (b)
DyN, and (c) TmN. The (i) refers to the rare earth site and the (ii) to the
nitrogen site. Full lines represent majority spin and dashed lines are 
minority spin} 
\label{fig4}
\end{figure*}

To discuss the properties of these materials more fully it is necessary
to understand the details of the density of states around $E_f$.
For this reason we show in Fig. \ref{fig4} the spin-resolved densities of states
on each site for SmN, DyN and TmN. These materials were chosen as typical
examples of the half-metallic, insulating and metallic rare earth nitrides
respectively. 
\subsection{The Light Rare Earth Nitrides}

We now explain how the rare earth nitrides can exhibit such a wide range of 
electronic properties in a series of materials that ostensibly have similar 
outer electronic structures. 
In  previous papers,\cite{nature,selen} we have 
calculated the electronic structure of rare earth chalcogenides. In 
agreement with observation and with the naive ionic model, we found that they 
are all divalent insulators. One might have expected the trivalent nitrides 
to be similarly insulating, but, as Table \ref{table2} shows, this is not the case. 
After the SIC-LSD has been applied to the rare earth nitrides 
the three valence electrons of the trivalent rare earth can fill the
three holes of the N p band and create a full $p$ band. This 
would lead to insulating behaviour in a similar manner that the two
electrons of the divalent rare earth in the rare earth chalcogenides 
fill the two holes in the chalcogenide $p$ band.  
In the nitrides there are three nitrogen $p$ electrons, in the 
chalcogenides there are four $p$ electrons. 
The difference in the number of $p$ electrons means that the $p$ bands sit closer to 
the Fermi energy in the nitrides than in the chalcogenides. In trivalent CeN 
the nitrogen $p$ bands are close enough to $E_f$ to hybridise slightly with the 
low energy tail of the rare earth $s$-$d$ bands, causing the materials to be 
metallic. As we proceed across the Periodic Table from CeN to GdN we fill the 
majority spin $f$-states and create an exchange field that is felt by the other 
electrons in the material. In the rare earth chalcogenides this has little 
effect because the filled bands are well down in the potential well. However, 
in the nitrides its effect is more pronounced and leads to a significant 
spin-splitting of the nitrogen $p$-bands. 

The SIC splits the $f$ band manifold into two, localized and band-like $f$ electrons.
The SIC $f$-bands are completely localised and are unable to hybridise significantly. 
The non-SIC $f$ states have a degree of itineracy that allows hybridisation. 
We find that N $p$-rare earth $f$ hybridised states play a role in the electronic bonding. 
This is clear in Fig. \ref{fig4}. In a(ii) panel for example, the 
N $p$-states, which lie within 0.2 Ry of $E_f$, lead to structure, in this energy
range, in the Sm density of states. This is an energy window where $f$ states do  
not occur, a $p$-$f$ hybrised bonding state has been formed. 
This occurs to some extent in all the rare earth nitrides.

Of course, there is also a $p$-$f$ antibonding state and it is this that is
key to understanding the trends in electronic properties in this series of
materials. The upper tail of the majority spin p-bands rises to above 
the Fermi energy and hybridises with the empty majority spin bands just above 
$E_f$ to form the antibonding state. For SmN this can be seen clearly in $a(i)$ 
and $a(ii)$ in Fig. \ref{fig4}. Just above $E_f$ the majority spin nitrogen $p$ 
density of states in $a(ii)$ has a small peak that corresponds with the rare 
earth $f$-bands in $a(i)$. 

The exchange field keeps the minority spin $f$-bands well above $E_f$ and 
hybridisation of them with the nitrogen $p$-bands is very small. In PrN and NdN 
the empty majority spin $f$-bands are well above $E_f$ and so the electronic 
properties are dominated by the majority spin rare earth $s$-$d$ bands. In PmN, 
SmN and EuN the empty majority spin $f$-bands have lowered and are very close 
to the Fermi energy. The density of states at $E_f$ is then dominated by $p$-$f$ 
bands. 

In the minority spin channel there is no significant $p-f$ hybridisation and 
so the predominantly $p$-bands are entirely filled and the $s$-$d$ bands are
empty. There is an energy gap at the Fermi energy. The minority spin bands
behave much more as expected in the naive ionic model.  

A consequence of this behaviour is that the nitrogen majority spin bands are
not fully occupied while the minority spin bands are. This means that the 
nitrogen moment is antiparallel to the rare earth moment. 

\subsection{The Heavy Rare Earth Nitrides}

In the heavy rare earth nitrides a repeat of the behaviour of the light
rare earths might be expected with the minority spin bands replacing the
majority spin ones. This does indeed occur. Inspection of Figs. \ref{fig4}b and
4c verifies that the comparable features can be seen in the minority spin
densities of states. However, other effects also come into play in the heavy
rare earth nitrides that modify the electronic properties significantly. 

In GdN, the majority spin $f$-bands are all filled and the minority spin 
bands are well above E$_{f}$. The exchange field continues to split the nitrogen
$p$-band. The majority spin $p$-band hybridises with the Gd $s-d$ bands. 
There is a minimum in the density of states very close to $E_f$, but it 
does not quite reach zero. This is very similar to what happens in CeN

In TbN, the minority spin $f$-bands begin to fill up. This pushes the majority
spin $f$-bands closer to the nucleus and so increases the efficiency of the
screening of the outer states so they are less strongly bound to the
nucleus. Because of this the rare earth $s-d$ states are higher above the 
Fermi energy in the heavy rare earths than in the light rare earths. In 
TbN, DyN and HoN then, there are no bands at the Fermi energy and therefore 
they are small gap insulators. In ErN, TmN and YbN the minority spin 
$f$-peak has come down to close to $E_f$ and there is a bonding and an 
antibonding minority spin state as in the light rare earths. As the $f$-peak 
comes down the antibonding state gets very close to $E_f$. The Fermi energy 
enters the major minority spin $f$-peak. To compensate for this some 
hybridised majority spin $p-d$ character has to rise above $E_f$ creating a 
non-zero density of states in both spin channels. 

The $p-f$ hybridisation pulls some minority spin character above the Fermi
energy and so for the heavy rare earths the nitrogen moment is parallel to 
the rare earth moment.    

For most rare earth nitrides very little experimental data appears to be
available. However, for YbN several papers have appeared in the literature
reporting experiments which can be compared with our work. Trivalency has
been established from the magnetic moment.\cite{Hull79} Ott {\it et
al.}, Sakon {\it et al.} and Takeda {\it et al.}\cite{Ott} reported 
measurements consistent with heavy fermion behaviour in YbN. However it 
was shown later by Monnier {\it et al.}\cite{Monn} that these results were 
also consistent with a scenario in which there is interplay between the crystal 
field and the Kondo effect for an isolated magnetic impurity. The SIC-LDA
method does not include the dynamical effects necessary to describe the
Kondo effect or heavy fermion behaviour, so we are unable to comment on this
point directly. However, we can point out that our results do exhibit the $p$-$f$ 
mixing necessary for the $p$-$f$ Kondo effect that explains the low temperature 
anomalies in the magnetic susceptibility and specific heat.\cite{Degi} 

The low temperature magnetic structure of YbN was established to be
antiferromagnetic III by D\"{o}nni {\it et al.}\cite{Donni}
Spectroscopic investigations of YbN include core level 
x-ray photoemission spectroscopy (XPS)\cite{Greber}
which revealed no evidence of valence mixing, and optical
spectroscopy\cite{Degi} which was able to locate the empty $f$ state 0.2 eV above 
the Fermi energy. They also found the bare occupied $f$-states to be around 
6.5 eV below the Fermi energy, which is not in good agreement with our
results, although there are always considerable uncertainties in locating
ground state bands using spectroscopic techniques.\cite{Pr}   
The XPS experiments also showed that samples exhibiting a small amount of 
non-stoichiometry can show strongly enhanced Kondo features. The experimental 
situation for YbN has been well summed up by Wachter\cite{Wach01} who shows 
that small amounts of doping could raise the Fermi energy to the empty $f$-state
in YbN and yield intermediate valence or heavy fermion behaviour. In our 
calculations the position of the empty $f$-state is in close agreement with the 
optical spectroscopy results and so we are also able to suggest that.
Furthermore the empty $f$-states in TmN are actually at the Fermi energy as
indicated in Table III. Extrapolating from the results for YbN suggests that
undoped TmN might exhibit heavy fermion or intermediate valent behaviour as
well. 

\section{Conclusions}

We have seen that the rare earth nitrides display a wide range of electronic
properties despite having the same crystal structure with only a small
variation in lattice constant and superficially similar electronic
structures. They show insulating (semiconducting), half-metallic and full
metallic behaviours. 
This is a consequence of the N $p$ and the band-like rare earth $f$ states
occuring in the same energy window, in the vicinity of the Fermi level.
This leads, at E$_{f}$, to strong hybridization of these states. 
With the high magnetic moments and the high
densities of states in one spin channel that these materials exhibit, it
suggests that alloying may enable us to fabricate materials with a wide and
continuous range of useful properties, particularly with regard to
spintronic and spin filtering applications.

\end{document}